\documentclass{PoS}
\usepackage{mkPhys}

\title{Tevatron Experimental Issues at High Luminosities}

\ShortTitle{Tevatron Experimental Issues at High Luminosities}

\author{\speaker{Michal Kreps}%
         \thanks{On behalf of the CDF and D\O{} collaborations.}
        \\Karlsruhe Institute for Technology\\
        E-mail: \email{michal.kreps@kit.edu}}

\abstract{
In this paper we describe the detector components, triggers
and analysis techniques for flavor physics at the Tevatron
experiments CDF and D\O{}. As Tevatron performs very well and runs at
higher luminosities regularly we also touch issues related
to it and efforts to improve detectors and triggers for such running.
}

\FullConference{12th International Conference on B-Physics at Hadron Machines - BEAUTY 2009\\
		 September 07 - 11 2009\\
		 Heidelberg, Germany}

\begin{document}

\section{Introduction}
The Tevatron $p\overline{p}$ collider with $\sqrt{s}=1.96$
TeV provides an excellent opportunity for the heavy flavor physics,
which is complementary to the $e^+e^-$ colliders operating at
$\Upsilon(4S)$ resonance. Main advantages are the large
$b$ hadron cross section of $\sigma=1.30\pm0.05\pm0.16$
$\mu$b for $b$ hadrons with $p_T>9$ \gevc and $|y|<0.6$
\cite{Aaltonen:2009xn},
which is three orders of magnitude higher than at $e^+e^-$
colliders and the large available energy for the production of
all sorts of $b$ hadrons. The main challenges consist of
a large inelastic cross section which is about three orders of
magnitude higher than $b$ hadron cross section and
a fragmentation at high energies, which produces a large amount
of particles in the detector. 
In this paper we discuss how we deal with the challenges,
starting from description of CDF and D\O{} detectors
and trigger strategies. For several important measurements,
flavor tagging is an important tool and we will describe its main
concept. Finally we discuss issues of running at
high luminosity and ways how we deal with them.

\section{Detectors}

Both CDF and D\O{} are multipurpose detectors exhibiting
cylindrical and forward/backward symmetry. Details of the
detectors are described in Ref.~\cite{Acosta:2004yw} for CDF and in
Ref.~\cite{Abazov:2005pn} for D\O{}. They have same basic structure
starting with  tracking detectors located in a magnetic field
in center, electromagnetic and hadronic calorimeters outside
the solenoid and a muon system behind the calorimeters.
In the following we use a coordinate system in which proton
beam direction defines $z$-axis. Angles
$\theta$ and $\phi$ are the polar and azimuthal angles
defined with respect to the $z$-axis. The pseudorapidity
$\eta$ is defined as $-\ln\tan(\theta/2)$ and the transverse
momentum is $p_T=p\sin(\theta)$. 

\begin{figure}
\centering
\includegraphics[width=0.45\textwidth]{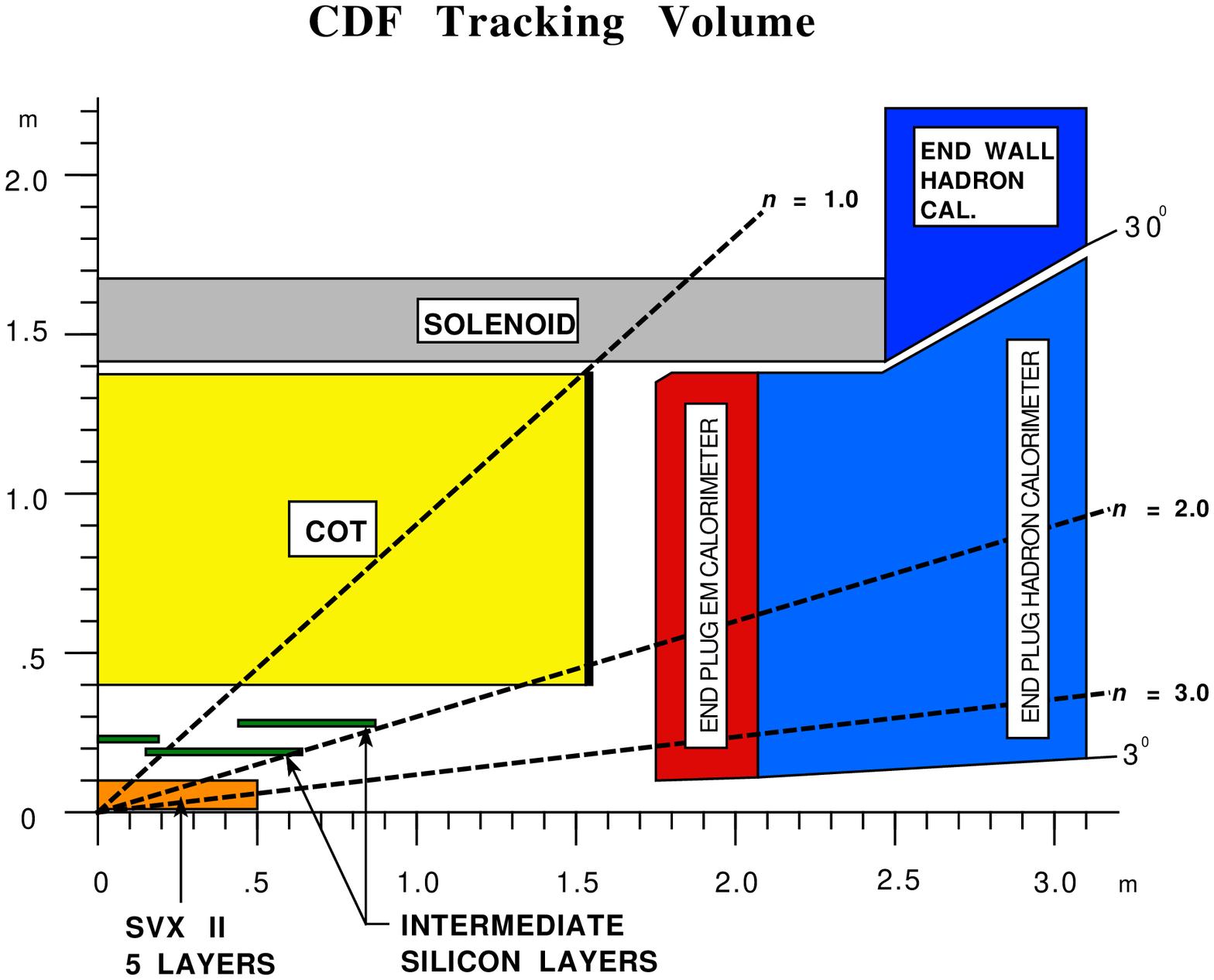}
\hfill
\includegraphics[angle=90,width=0.45\textwidth]{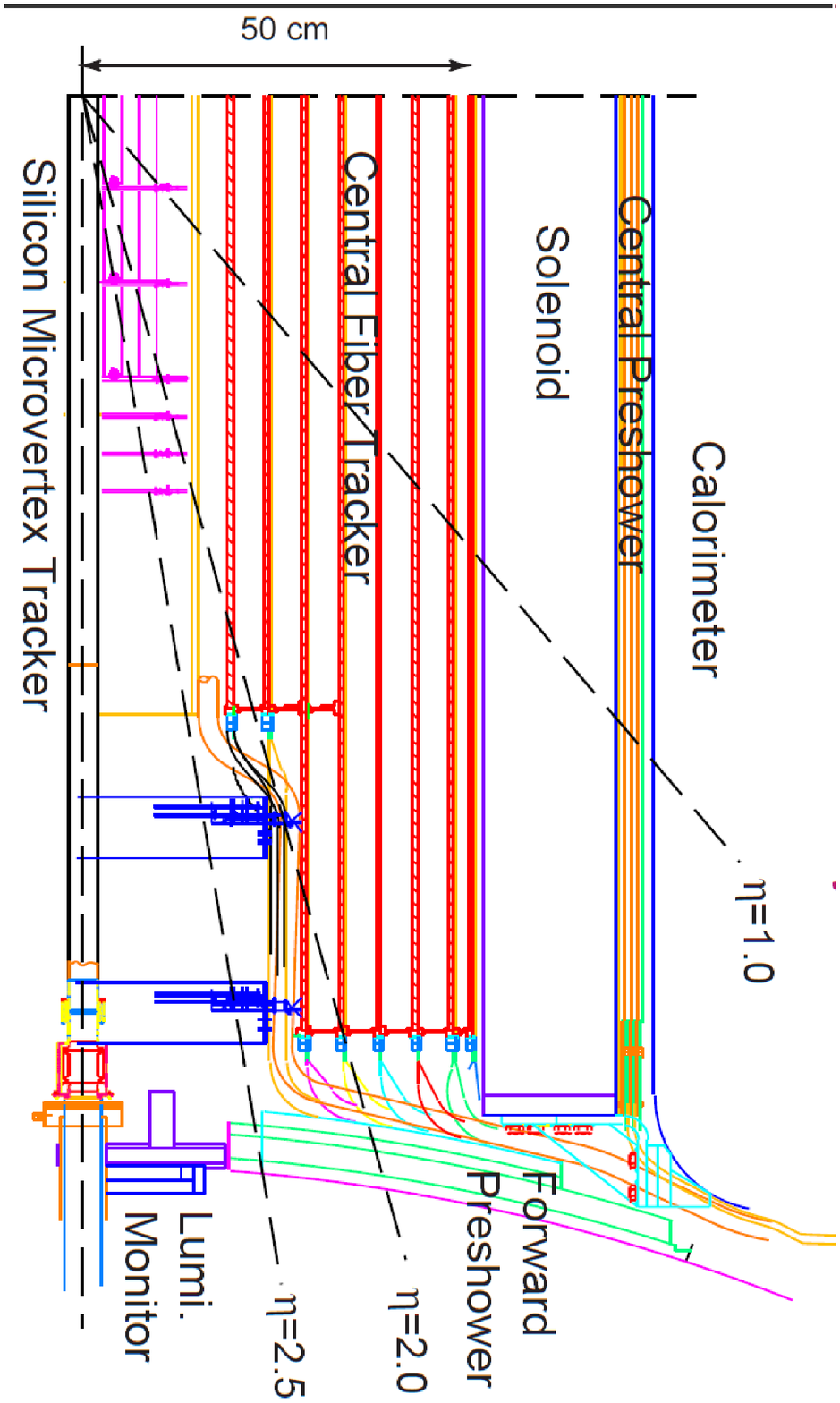}
\caption{Schematic view of the CDF (left) and D\O{} (right)
trackers.}
\label{trackers}
\end{figure}

The CDF tracking consists of silicon tracker close to
the interaction point, which is complemented by a large volume open cell
drift chamber (COT), both immersed in the 1.4 T magnetic field.
Schematic view of the tracking system is shown in
Fig.~\ref{trackers}. The silicon detector itself
consists of three subcomponents. First one is a single layer of
one sided silicon strip detector (Layer00) mounted directly on beam
pipe at a distance of 1.5 cm from the beamline
\cite{Hill:2004qb}. The main part consists of
5 layers of double sided strip detector (SVX II) with radial
extent between 2.5 and 10.6 cm from the beamline
\cite{Sill:2000zz}.
The last part called ISL provides single double sided layer for
$|\eta|<1$ at radius of 22 cm and two layers at 20 and 28 cm for
$1<|\eta|<2$ \cite{Affolder:2000tj}. The active volume of
the COT extents from 43.4 to 132.3 cm from the beamline and
provides a full coverage
for tracks with $|\eta|<1$ \cite{Affolder:2003ep}. 
Cells are grouped into 8 superlayers in radial direction, each consisting of 12
cells. Four of the superlayers are axial and four have small
stereo angle to provide information in the $r-z$ view. The
two types of superlayers alternate with outermost one being
axial superlayer. The integrated tracker provides a
transverse momentum resolution of $\sigma(p_T)/p_T^2=0.15\%
(\mathrm{GeV}/c)^{-1}$.

The D\O{} tracking consists of a silicon microstrip detector
complemented by a fiber tracker at larger distances from the
beamline, both immersed in the 2.0 T magnetic
field. Layout of the tracking system is shown if
Fig.~\ref{trackers}. The silicon detector provides four layers of double
sided sensors in the central region and two double sided with
two single sided sensors in the forward/backward regions. The
innermost layer is located at 2.7 cm and outermost at 10.5
cm. In the forward and backward region additional disks of
silicon sensors are installed perpendicularly to the beam
direction to further enhance tracking at large $|\eta|$.
In 2006 the silicon detector was extended by an additional
layer mounted close to the beampipe in order to improve
the impact parameter resolution of tracks. The fiber tracker
is based on scintillating fibers. The 16 layers are located
from radius of 20 cm up to radius of 52 cm. The full
tracking detector provides coverage up to $|\eta|<2$.

Both experiments contain electromagnetic and hadronic
calorimeters covering large part of the phase space. 
They are optimized for high $p_T$ physics and thus only of
limited use in heavy flavor physics. 
The electromagnetic calorimeters are used for electron identification with main
use in  flavor tagging. Photon detection is possible to some
extent, but none of the experiments exploited it up to now.

Muon detection is one of the key components for flavor
physics at Tevatron. CDF detects muons in a set of drift
chambers and scintillators. The central muon detector
\cite{Ascoli:1987av} is located outside the hadron calorimeter and
covers $|\eta|<0.6$. In the same central region 
detectors behind an additional absorber provide improved
identification in exchange of increased $p_T$ threshold. For
run 2 the muon coverage was extended up to $|\eta|<1.0$.  
The $p_T$ threshold of the muon systems is between $1.5$ to
$2.0$ \gevc depending on the $|\eta|$ region.
The muon system of D\O{} consists of drift tubes complemented
by scintillator counters. The main difference to the CDF system
is in coverage, where D\O{} system covers up to
$|\eta|<2.0$. Thanks to an additional toroid with magnetic field of
1.8 T, the D\O{} muon system provides
standalone momentum measurement for muons. In addition,
there is a more material in front of the muon chambers, which
results in a lower background compared to CDF.

\begin{figure}
\begin{minipage}[c]{0.47\textwidth}
\centering
\includegraphics[width=0.9\textwidth]{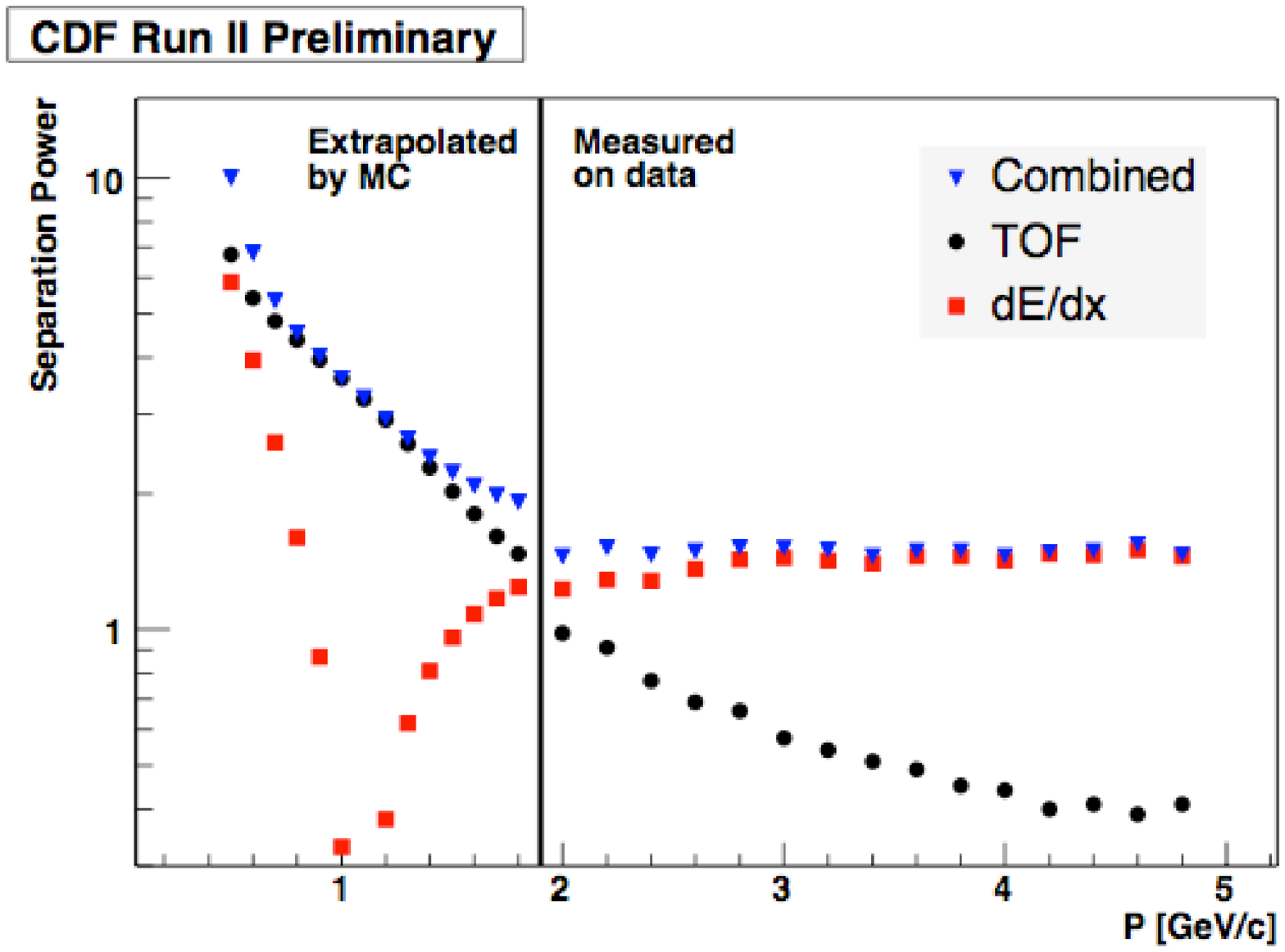}
\caption{Separation power of CDF particle identification
system. The black points show TOF, the red points
$\mathrm{d}E/\mathrm{d}x$ and the blue points likelihood
ratio combination of the two.}
\label{CDFPID}
\end{minipage}
\hfill
\begin{minipage}[c]{0.47\textwidth}
\centering
\includegraphics[angle=90,width=0.9\textwidth]{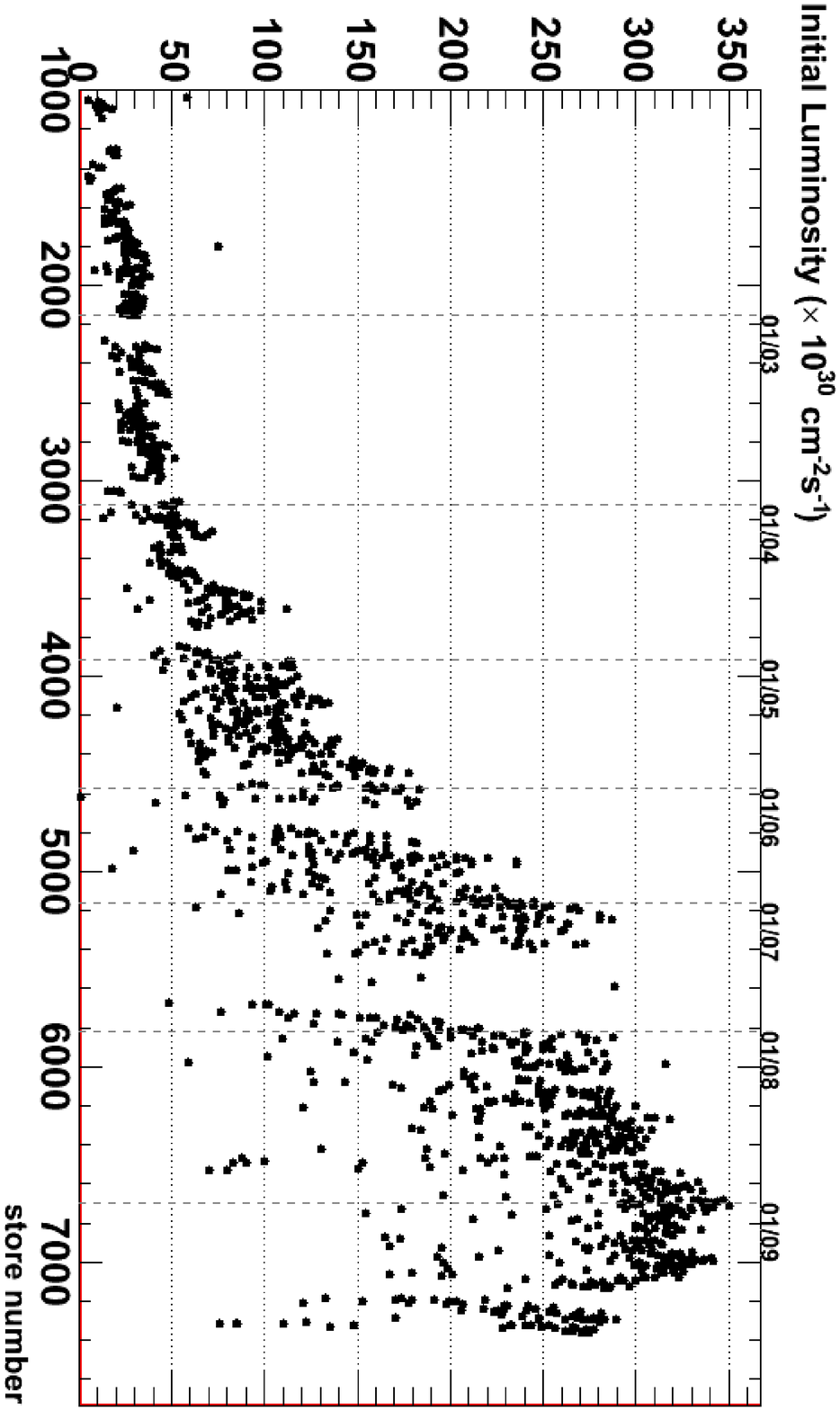}
\caption{Initial luminosity of Tevatron accelerator as a
function of time.}
\label{initLum}
\end{minipage}
\end{figure}
While D\O{} does not have dedicated particle identification
for hadrons, CDF experiment utilizes two detectors to
separate kaons from pions. The first information is provided by
the COT, where thanks to its large volume and number
of measurements along the track, $\mathrm{d}E/\mathrm{d}x$
measurement is used for separation. This is complemented
by the time-of-flight detector (TOF) \cite{Acosta:2004kc},
which is located between  COT and solenoid. It consists of 216 Bicron
scintillator bars and covers $|\eta|<1.0$. With time
resolution on the single hit of 110 ps it provides
better than $2\sigma$ separation between kaons and pions
for $p_T<1.5$ \gevc. Figure \ref{CDFPID} shows the separation
power as a function of momentum for the combined system. The
separation power is at least $1.5\sigma$ over the whole
interesting momentum region. It is worth to note that
$\mathrm{d}E/\mathrm{d}x$ and TOF measurements complement
each other. While separation is not sufficient for
track-by-track identification, it provides reasonable
statistical separation.

To summarize detectors, both provide reasonable performance
for flavor physics at hadron colliders and both experiments
contributed with important measurements. D\O{} detector has
its strengths in the muon detection with considerably larger
coverage in $|\eta|$. Additional advantage not discussed here
is in a regular swap of the polarity of magnetic field in
solenoid and toroid magnets. On contrary, CDF
detector thanks to the larger radial extent of the tracker has
excellent momentum resolution, which reflects itself to a very
good mass resolution. The particle identification provides
additional advantage over D\O{} detector.

\section{Trigger strategies for $b$-physics}
While the $b$ hadron cross section is rather large, the
inelastic cross section is three orders of magnitude larger.
With a collision rate of 1.7 MHz and a size of an
event of order 100 kB the data flow is of order 200 GB per
second. Such data rate is beyond current storage system
capabilities. Therefore flavor physics at hadron colliders
critically depends on the ability of the trigger to select  heavy
flavor events with good efficiency and huge background
rejection.
The basic three level architecture of the trigger system is
the same in both
experiments. The level 1 is implemented purely in hardware
and the main information used for its decision comes from the muon system
and the tracking in the fiber tracker at D\O{} and the COT
at CDF. The level 2 is a mixture of hardware and software
implementation, which allows  more flexibility. Here both
experiments have access to the whole detector including
silicon tracker, but only CDF exploits silicon information
for triggering. Finally, level 3 is implemented purely
in software running on dedicated computing farms. It 
runs a speed optimized version of the offline reconstruction
software. In most of the cases, the level 3 reconfirms
decisions of the previous levels with higher precision. 
The D\O{} level 1 trigger has latency of 4.2 $\mu$s and
reduces the accept rate from 1.7 MHz to about 2.5 kHZ. Level 2 has
about 100 $\mu$s for its decision and reduces the rate to
approximately 1 kHz while the level 3 
reduces the rate to a final value of 200 Hz. At  CDF the level 1 has about 5.5
$\mu$s for its decision and reduces the rate down to about 20-30
kHz. The level 2 has an average latency of 20 $\mu$s with
an output rate of about 700 Hz. Finally level 3 provides an output
rate of about 120 Hz. 

The trigger strategy at D\O{} is based purely on  muon
detection. There are two classes of triggers, one requiring
two muons of opposite charge and one with single muon
requirement. First strategy is used for studies of decays
with \Jpsi in the final state like the $\beta_s$ measurement
\cite{Abazov:2008fj} 
or searches for rare decays with prime example 
$\Bs\rightarrow \mu^+\mu^-$ \cite{Abazov:2007iy}. 
The second class of
triggers is used with semileptonic decays (e.g.
study of $\Bs\rightarrow D_s^{(*)+}D_s^{(*)-}$ with one
$D_s$ decaying semileptonically) or fully hadronic decays
used in the \Bs mixing measurement \cite{Abazov:2008mix}. 
The triggering of fully hadronic decays is based on the
assumption that the muon originates from
a decay of the second $B$
hadron in the event. The triggered sample has lot of
background, but every single event can be flavor tagged
using the trigger muon.

At CDF thanks to the use of information from silicon detector
the trigger strategy is different from D\O{}. The main information at
level 1 comes from the muon detection and use of the extremely
fast tracker (XFT) \cite{Thomson:2002xp} which finds tracks
in the axial layers
of the drift chamber. The $p_T$ threshold of the XFT is 2 \gevc.
At level 2 the silicon vertex trigger (SVT)
\cite{Ashmanskas:2003gf} play a
crucial role in adding hits in silicon to the tracks found
by XFT and in measuring impact parameters of tracks with
a typical resolution of order 50 $\mu$m including the
beam size. This allows for the selection of tracks coming from displaced
vertices and thus triggering also on decays which do not
provide lepton in final state. Altogether, three classes of
triggers are used. First class requiring dimuon pair of
opposite charge is similar to D\O{} and is used for similar
physics like measurement of $\beta_s$ \cite{Aaltonen:2007he}
or search for $\Bs\rightarrow \mu^+\mu^-$
\cite{Aaltonen:2007kv}. The second class requires one lepton (muon or
electron) with $p_T>4$ \gevc and an additional track with large
impact parameter. This class is designed for semileptonic $B$
decays. The third class requires two tracks with a large
impact parameter (at least 100 $\mu$m) which are consistent with common vertex
with a sizeable displacement from the beamline and CDF is
first experiment to implement this kind of trigger. 
It is this last trigger strategy that allows to select
decays like $\Bs\rightarrow D_s\pi$ used for the \Bs mixing
observation \cite{Abulencia:2006ze} or $D^{*+}\rightarrow D^0\pi^+$ which
provides largest samples in the world for $D^0$ mixing and CP violation
measurements \cite{Aaltonen:2007uc}.

\section{Flavor tagging}
One of the important tools for the physics goals of the
Tevatron run 2 is a flavor tagging of the $B$ hadrons. The
task of the flavor tagging is to find out whether a $B$ hadron
was produced as $B$ or $\overline{B}$. Without such tool the \Bs
mixing \cite{Abulencia:2006ze,Abazov:2008mix} would be impossible. It is also crucial tool
for the time dependent CP violation measurements, which
experiments try to perform these days.

At Tevatron, $b$ quarks are produced in pairs of opposite
flavor. After hadronization, they end up in different $B$
hadrons. One which is used in a given study is usually called
same side $B$ hadron. The other $B$ hadron is often referred
to as the opposite side $B$. The flavor tagging is then split into
two categories, the opposite side taggers (OST) and the same
side tagger (SST). The OST exploits decay of the opposite
side $B$, while SST examines fragmentation
partners of the same side $B$. 
The performance of the flavor tagging is characterized by
efficiency $\epsilon=N_{tag}/N_{tot}$ which a gives fraction
of events in which a decision is made and dilution
$D=(N_{correct}-n_{wrong})/N_{tag}$ which is related to the
probability $P$ that event is correctly tagged by $D=2P-1$.
The quantity $\epsilon D^2$ than defines an effective statistic
corresponding to perfectly tagged events.
Some details of the
algorithms are described in Ref.~\cite{Moulik:2007px}.

The OST tagging is based on the leptons arising from
semileptonic decays, jet charge or secondary vertex charge
and charge of the kaon from the decay chain $b\rightarrow
c\rightarrow s$. We use both muons and
electrons for the lepton tagging. At CDF, a likelihood based
selection is used while D\O{} uses rectangular cuts for
lepton identification. In CDF jet charge tagger is employed,
while D\O{} uses secondary vertex charge. The jet charge
tagging of CDF starts with track based jets in which it
tries to search for a secondary vertex. For each jet, charge
is calculated as $Q_{jet}=\sum q_ip_T^i/\sum p_T^i$, where
sums run over all tracks associated with jet. CDF, thanks to
its capability of $K/\pi$ separation, is able to exploit also
opposite side kaon tagging. In case of both experiments,
different taggers are combined using multivariate techniques
to a single opposite side decision. Altogether, as D\O{} has
the better muon coverage and cleaner muon detection, they
achieve better performance than CDF.

The SST is based on the fact that to produce \Bs one needs
an $s$ quark from $s\overline{s}$ popup. Afterwards
$\overline{s}$ quark has to end up in a hadron, which is often
$K$. If it is a charged kaon, its charge defines, whether it
contains quark or anti-quark and thus also defines
production flavor of the \Bs. Both experiments exploits this
idea for flavor tagging. While D\O{} is selecting a tagging
track (fragmentation partner) based on the kinematical
relation of tracks to a \Bs candidate, CDF combines
kinematical information with particle identification
information for this task. Thanks to particle
identification, CDF has better performance in this tagger.

Both experiments have tagging power of $\epsilon D^2\approx
4.5\%$. The opposite side taggers are usually calibrated by
measuring well known $B^0$ oscillations as it exploits
information which is independent of the same side $B$ hadron.
The SST is developed and studied on simulated events, but
\Bs mixing measurements indicate that understanding of the
taggers is reasonable. Unfortunately the \Bs mixing measurement itself
did not provide sufficient statistics for a high precision
calibration up to now.

\section{High luminosity running and upgrades}
Tevatron went through many optimizations and improvements in operation
procedures over the past two years \cite{Convery:2009zq}. Those
improvements result in a better stability of the accelerator
as well as in  higher initial luminosities for stores. In
Fig.~\ref{initLum} we show development of the initial
luminosities as a function of time. It can be seen that
since about mid 2008 they are consistently above $300\cdot
10^{30}$ cm$^{-2}$s$^{-1}$.
\begin{figure}
\end{figure}

\begin{figure}
\centering
\begin{minipage}[c]{0.40\textwidth}
\includegraphics[width=0.75\textwidth,angle=90]{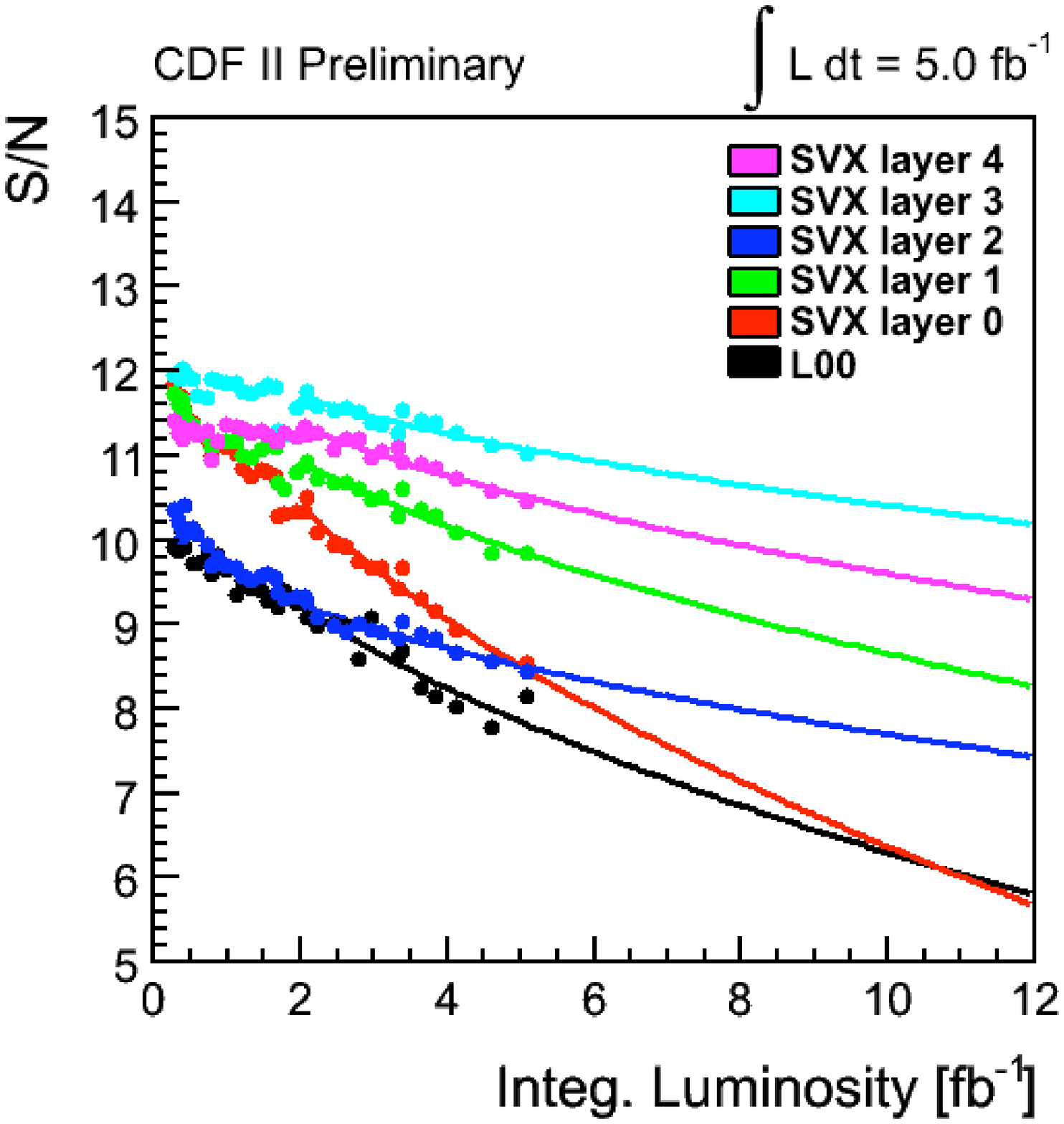}
\end{minipage}
\begin{minipage}[c]{0.50\textwidth}
\includegraphics[width=0.75\textwidth]{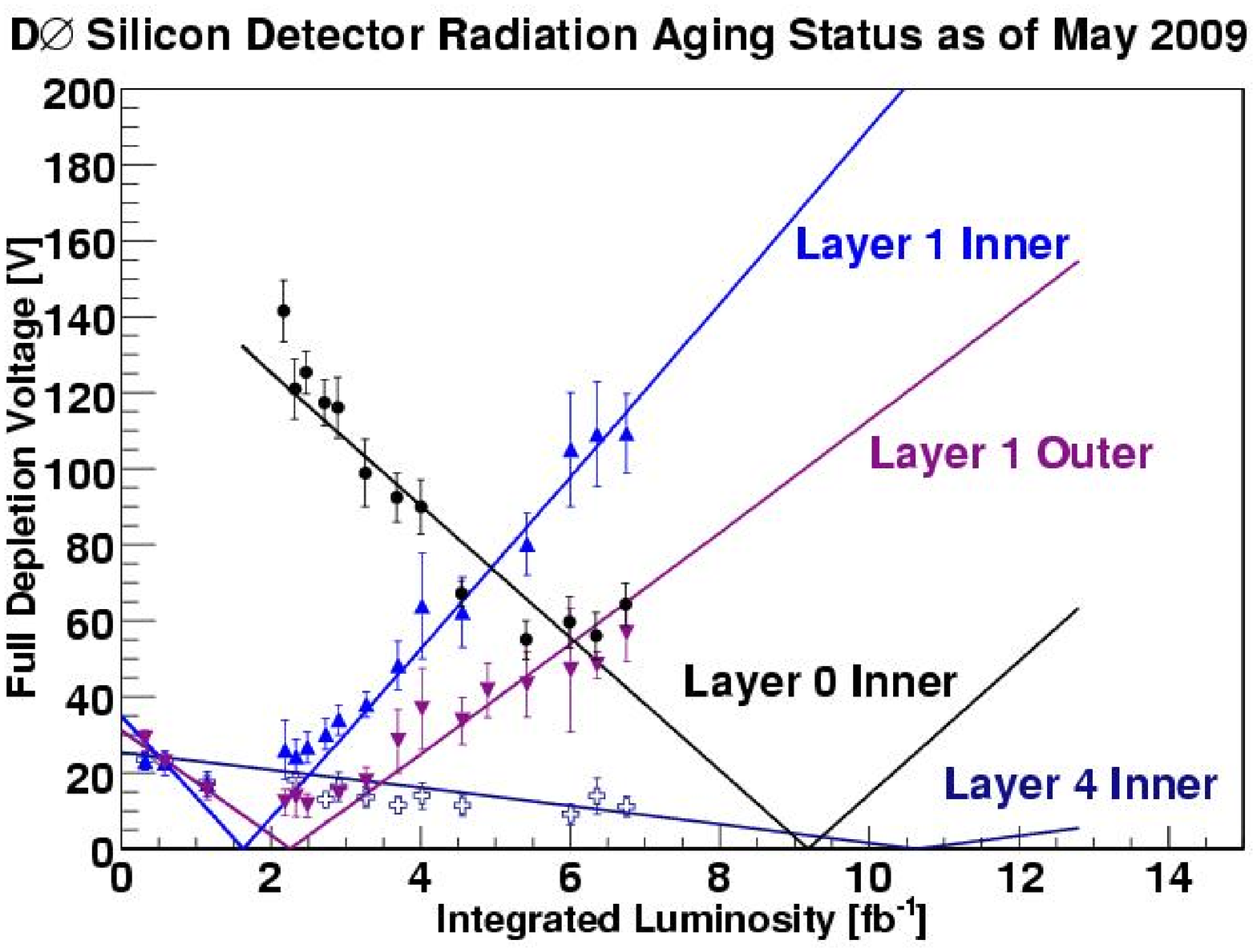}
\end{minipage}
\caption{The signal to noise ratio for the CDF silicon
detector (left) and the depletion voltage of the D\O{} silicon
detector (right) as a function of luminosity.}
\label{siliconStatus}
\end{figure}

As the luminosity increases, demands on the detectors and
trigger systems increase as well. This is due to the fact
that at luminosities above $300\cdot 10^{30}$
cm$^{-2}$s$^{-1}$ we get on average about 10 inelastic
$p\overline{p}$ collisions. It results in more particles
inside the detector and thus more information which trigger
system needs to process and higher chance of the coincidence
of unrelated trigger objects to pass the trigger requirements. Additionally
also radiation seen by the detectors increases,
which is specially important for the sensitive silicon
detectors, which are mounted closest to beamlime.
In Fig.~\ref{siliconStatus} we show the signal to noise 
ratio for CDF silicon detectors and the depletion voltage of the
D\O{} detector as a function of luminosity.
Currently we collected about 6 \invfb of data and silicon
detectors of both experiments are fully operational. For CDF
we expect no significant degradation in offline performance
for $S/N$ above 3 and in trigger performance for $S/N$ above
6. As can be seen from Fig.~\ref{siliconStatus} the silicon
detector is expected to perform without significant
degradation up to about 12 \invfb. At D\O{}, detector
projection tell that the silicon detector can be operated
without large issues up to 12 \invfb. Only significant issue
might be with the Layer 1, which is mitigated by installation of
the Layer 0 in 2006. No significant performance degradation is
therefore expected even if Layer 1 stops to operate. 
From the offline tracking point of view additional interactions
per bunch crossing do not cause any significant issues. This is
given by the length of the bunches, which is large enough to
provide primary vertices with large enough separation in
order to distinguish them by tracking.

\begin{figure}
\centering
\includegraphics[width=0.32\textwidth]{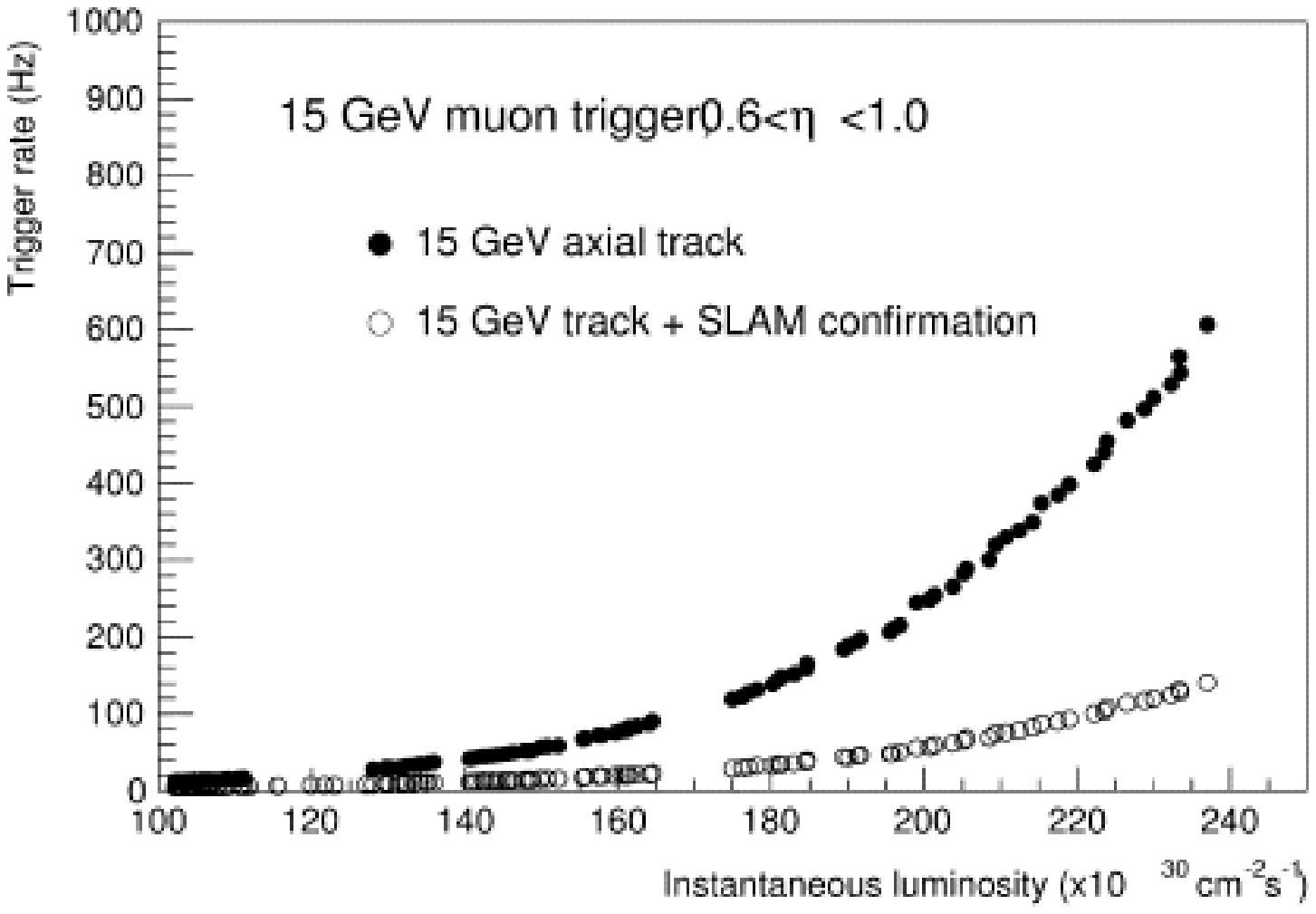}
\hspace{0.1\textwidth}
\includegraphics[width=0.28\textwidth]{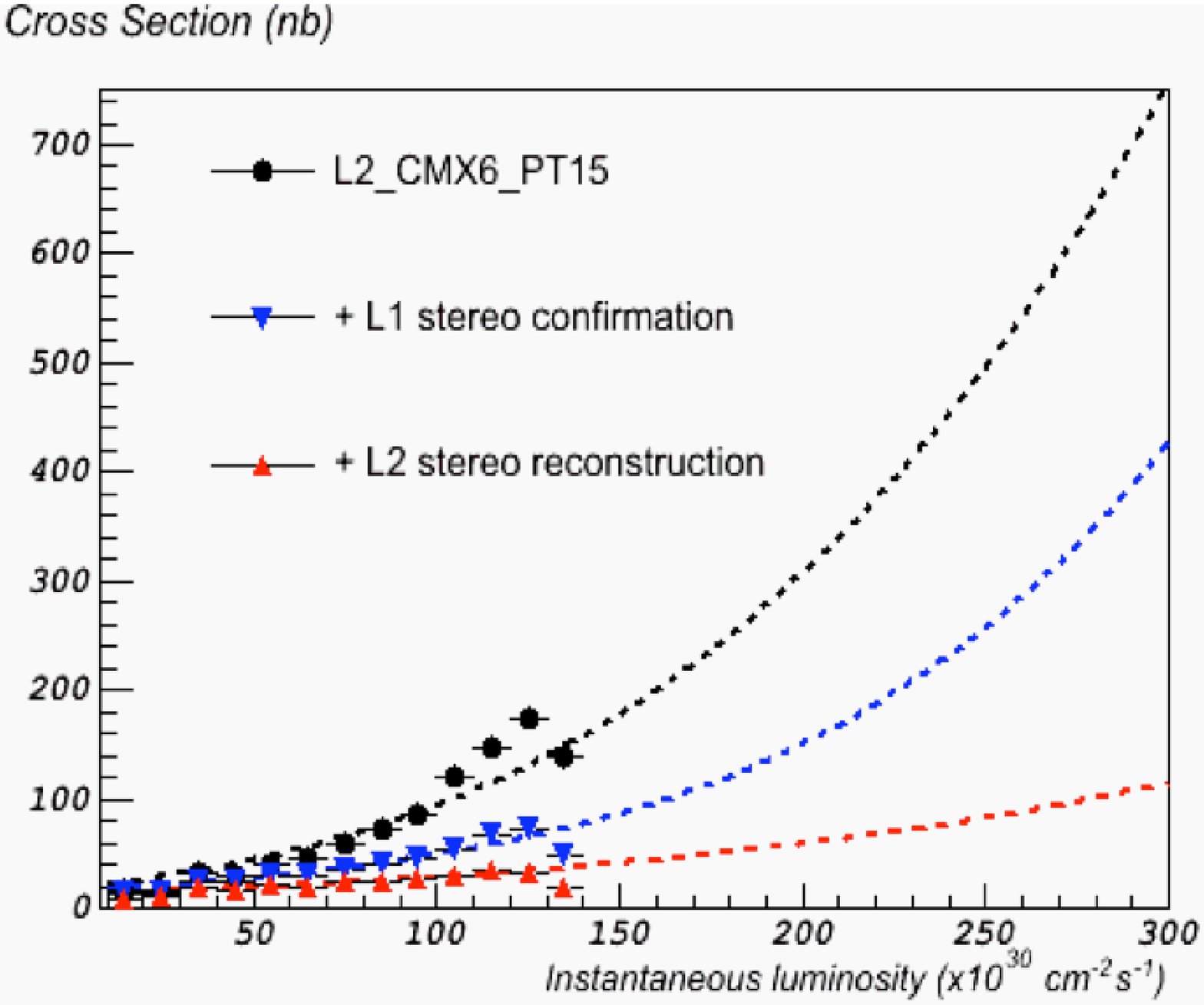}
\caption{Example of the effect of the XFT upgrade which adds
stereo tracking on the trigger rate at level 1 (left) and at
level 2 (right). Example shows single high $p_T$ muon trigger,
not directly used by heavy flavor physics.}
\label{XFTupgrade}
\end{figure}

The main issue related to the high luminosity running comes
with trigger. Trigger rates often increase more than
linearly as one would expect from a pure physics of interest.
Continuous effort is made by experiments to optimize their
trigger strategies to keep up with the performance of the
accelerator. As both CDF and D\O{} are multipurpose
experiments, it sometimes results in optimization of not
only heavy flavor triggers, but also high $p_T$ triggers.
At CDF we perform additional upgrades of the trigger systems
to cope with higher rates. First of the upgrades, which is
in successful operation was the upgrade of XFT system
\cite{Abulencia:2007xft}. Original
system was capable to perform tracking only in axial layers
of the COT. The upgrade added possibility of tracking also
in stereo layers, which improves the background rejection
significantly. Example of the effect of the upgrade can be
seen on Fig.~\ref{XFTupgrade}.
Second upgrade currently ongoing concerns the SVT subsystem.
Already in 2006 we exchanged Associative Memory boards to ones with more
memory which are capable to
store more patterns. This upgrade allowed to improve
resolution and decrease the background rate. Currently we are in
a commissioning phase of a second upgrade which exchanges track
fitting boards by GigaFitter board, which is a new generation of
online track fitting processor \cite{Amerio:2007giga}. It allows further
increase in precision as well as removes prolongations in
the processing time for events with more tracks.

\section{Conclusions}
To conclude, both Tevatron experiments are able to provide
important results in the heavy flavor sector, including
highly nontrivial measurements of the time dependent CP
violation and neutral meson mixing. Both collaborations are
still in the position to keep up with improvements on
the accelerator side and thanks to the effort of accelerator and
detector experts we are collecting new data with
an unprecedent rate. With approved running in the fiscal year
2010 we expect to collect about 8 \invfb by each of the
experiments. In likely case of running through 2011 we
expect to have about 10 \invfb of good data available for analysis at
each of the experiments. 
Altogether, there is strong will inside
the community to provide over next few years competitor in
the physics to the LHC experiments.

\end{document}